\documentclass[
aps
pre,
twocolumn,
nofootinbib,superscriptaddress,showkeys,showpacs,
preprintnumbers, 
amsmath,amssymb,final]{revtex4-2}

\usepackage{graphicx,color}
\usepackage{dcolumn}
\usepackage{bm}
\usepackage{graphicx}
\usepackage{dcolumn}
\usepackage{bm}
\usepackage{color}

\usepackage{subfigure}

\usepackage{mathptmx}  
\usepackage{setspace}  

\usepackage{orcidlink}

\usepackage{soul}
\usepackage{tikz}

\usepackage{hyperref}
\hypersetup{colorlinks=true,linkcolor=blue,urlcolor=blue,citecolor=blue}
\usepackage[toc,page]{appendix}
\usepackage[normalem]{ulem}

\hypersetup{colorlinks = true,linkcolor = blue,anchorcolor = blue,citecolor = blue, filecolor = blue,urlcolor = blue}
\usetikzlibrary{shapes,arrows,positioning,fit,backgrounds,calc}
\usetikzlibrary{shapes.geometric, arrows}
\usepackage{hyperref}
\usepackage[mathlines]{lineno}
\begin{document}

\preprint{APS/123-QED}

\title{A New Model for the Income Distribution} %

\author{Ekrem Aydiner}

\email{ekrem.aydiner@istanbul.edu.tr}


\affiliation{Department of Physics, Faculty of Science, İstanbul University, İstanbul 34134, Türkiye}

\thanks{This work was carried out during a visit to Princeton University.} 

\date{October 14, 2025, Princeton}%

\begin{abstract}

In this study, we propose a kinetic trap--diffusion model to describe the
emergence of Pareto distributions in money-exchange systems.
Using kinetic Monte Carlo simulations, we show that the Pareto exponent depends
explicitly on temperature and takes values in the range
$0.5 \leq \nu(T) \leq 1.5$.
In the present framework, the temperature $T$ acts as a control
parameter that regulates the exchange dynamics through thermally activated
diffusion.
Unlike conventional kinetic exchange models, where the Pareto exponent is fixed
by microscopic rules, the proposed model generates a range of Pareto exponents
as a function of $T$.
The temperature-dependent Pareto exponent constitutes the central novelty of
the proposed kinetic trap--diffusion framework.
This feature provides a natural explanation for the empirically observed
variations in Pareto exponents across different countries and economic
conditions.
In addition, the fraction of zero-wealth agents and the Gini index
exhibit a non-monotonic dependence on temperature, revealing distinct
dynamical regimes arising from the competition between trapping and
money mobility. The persistence of the Pareto-like stationary
distribution under strongly nonuniform initial conditions further
supports the robustness of the proposed mechanism. These results show
that different Pareto-tail and inequality regimes can emerge from the same
trap--diffusion dynamics through changes in a single control parameter.

\end{abstract}

\keywords{Pareto law, income distribution, money distribution}
\maketitle


\maketitle


\section{Introduction} \label{intro}

The distribution of income or wealth in a society or a country often exhibits
a Pareto-type behavior~\cite{Pareto1897}, which is commonly used to quantify
the degree of inequality~\cite{gini21,elia12gini-soko,elia16harness}.
As first observed empirically by Pareto in 1897, the upper tail of the income
distribution $P(x)$ follows a power-law form
\cite{Pareto1897,mand60pareto-levy,bouc00conden-wealth,pers92}
\begin{equation}
P(x) \sim x^{-\nu},
\label{eq-first-pareto-distr}
\end{equation}
where the scaling exponent is typically found to take values of order unity.
In particular, analyses of income data suggest that the exponent
characterizing the upper tail is often reported to be close to
$\nu \approx 3/2$ in cumulative representations, although its precise
value depends on the dataset and on the adopted definition of the
distribution.
Remarkably similar scaling exponents were observed by Pareto across different
societies, including Italian cities, England, several German states, Prussia,
as well as historical data from ancient Peru and the Cherokee Indians~\cite{pers92}.
Over the past century, the Pareto exponent has been found to vary only
moderately in time and across different capitalist economies
(see, e.g., Ref.~\cite{kato22islam-capital} for comparisons with Islamic
economies).
Gibrat later clarified that the Pareto law applies exclusively to the
high-income range~\cite{Gibrat1931}.
More detailed empirical studies have shown that income distributions are well
described by a log-normal form for small incomes ($x \to 0$), while exhibiting
a Pareto-type power-law tail at large income values
\cite{Dragulescu_2001,Fujiwara2003}.

In recent years, empirical and theoretical investigations of inequality across
different societies have flourished~
\cite{gaba09rev-pareto-power-laws, yako10njp, saez14inequal-science,
levi14big-data, elia14rich-poor, chak16inequality-kin-exch,
gaba16inequality-well-cited, shes17unequal-psycho,
diff19glob-warm-pnas-inequal, stoj22rgbm-income, kemp22}.
Analyses based on large-scale income and wealth datasets from the United States,
Japan, and European economies consistently confirm the presence of Pareto-type
power-law tails. Similarly,  studies of personal income distributions in the United States, Japan, and European
economies demonstrate that the exponent typically lies in the range
$1 < \nu < 3$, reflecting differences in economic inequality, taxation,
and market structure~
\cite{Dragulescu_2001, Souma_2002, Fujiwara_2003,
Atkinson_2011, Clementi_2005}.
These findings confirm that power-law behavior is a robust feature of real
economies, while the precise value of the Pareto exponent is not universal but
depends sensitively on macroeconomic and institutional factors that shape wealth
inequality.

It can be seen in the literature that a large number of models have been proposed to describe and explain the statistical properties of money distributions observed in real economies. 
These models differ in their underlying mechanisms, ranging from microscopic exchange processes to macroscopic growth dynamics, and can be broadly categorized according to their fundamental principles
\cite{Dragulescu2000,Chakraborti_2000,Chatterjee2005,Champernowne1953,Garlaschelli2004,Atkinson2011,Yakovenko_2009}.

Among these models, in recent years, \emph{kinetic exchange models} have attracted increasing attention in studies aiming to explain the distribution of income and wealth/money \cite{Dragulescu2000,Chakraborti_2000,Chatterjee2005,Champernowne1953,Garlaschelli2004,Atkinson2011,Yakovenko_2009,yako01usa-income,Boghosian_2014,Cordier_2005,chak13book-income-inequal-wealth,chak22income-distr-inequal,Silva2005,Chakraborti_2002, Chatterjee_2003,Chatterjee_2004,Chatterjee_2007}. These models represent one of the most important theoretical frameworks developed for understanding wealth and income inequality. Inspired by the kinetic theory of gases, they describe monetary exchanges between economic agents as analogous to energy exchanges between gas molecules, while assuming that the total amount of money in the system is conserved. As a result of repeated pairwise interactions, the system evolves toward a stationary money distribution similar to the Boltzmann--Gibbs form. When heterogeneous saving propensities are introduced among agents, the steady-state distribution develops Pareto-type heavy tails. Foundational contributions to this approach include the pioneering works of Dr\u{a}gulescu and Yakovenko~\cite{Dragulescu2000,yako01usa-income,Silva2005,Yakovenko_2009}, Chakraborti and Chakrabarti~\cite{Chakraborti_2000,Chakraborti_2002,Chatterjee_2007}, Chatterjee, Chakrabarti, and Manna~\cite{Chatterjee_2003,Chatterjee_2004}. 
Risk-based and stochastic extensions of kinetic exchange models, such as the Yard--Sale and Cordier--Pareschi--Toscani formulations, have also been shown to generate exponential or Pareto-type wealth distributions \cite{Boghosian_2014,Cordier_2005}.

Despite their success, such models typically generate a single Pareto exponent.
In contrast, different countries and economic systems are characterized by
distinct Pareto exponents.
This indicates that income and wealth inequality cannot be described by a
universal and fixed scaling exponent.
Therefore, models incorporating mechanisms capable of producing different
Pareto exponents are of particular importance for a realistic description of
economic systems.

In this work, we propose a kinetic trap--diffusion model of money exchange that
extends standard kinetic exchange approaches by introducing temperature as a
 control parameter.
In the proposed framework, money is exchanged between agents according to a
temperature-dependent diffusion dynamics, while the total amount of money is
conserved.
As a result, the effective mobility of money varies with temperature, directly
affecting the exploration of the underlying potential landscape formed by
heterogeneous traps.
When the model is simulated at different temperatures, distinct stationary
wealth distributions emerge, characterized by different Pareto exponents in
the high-income regime.
Hence, unlike conventional kinetic exchange models that produce a single,
fixed Pareto exponent, the present model naturally generates a continuous
family of Pareto exponents as a function of temperature, providing a
physically motivated explanation for the diversity of inequality patterns
observed across real economies. In addition, the vacancy fraction and the Gini index exhibit
non-monotonic temperature dependence, indicating distinct dynamical
regimes associated with the competition between trapping and money
mobility. The persistence of the Pareto-like behavior under different
initial wealth configurations further demonstrates the robustness of
the proposed mechanism.

This work is organized as follows. In Section~\ref{KineticEM}, we briefly summarize kinetic exchange models. Section~\ref{KineticTDM} introduces the kinetic trap--diffusion model. Monte Carlo simulation results are presented in Section~\ref{MCSim}. A detailed discussion of the results is given in Section~\ref{Disc}. Finally, the main conclusions are summarized in Section~\ref{Conc}.

\section{Kinetic Exchange Models} \label{KineticEM}

A statistical model of a \emph{closed money-exchange system}---analogous to the kinetic model of an ideal gas---can be solved exactly or simulated numerically. Here, $N$ is the number of the agents,  $M$ denotes total money in the system. The state of the $i$-th agent is characterized by $x_i$, where $i = \{1,2,\ldots,N\}$, and the total amount of money is conserved:
\begin{equation} \label{Constant_M}
M = \sum_{i=1}^{N} x_{i} = \text{const}.
\end{equation}

At each time step, two randomly chosen agents $i$ and $j$ exchange an amount $\Delta x_{ij}$:
\begin{eqnarray} \label{delta_m}
x_{i}^{\prime} = x_{i} + \Delta x_{ij}, \qquad
x_{j}^{\prime} = x_{j} - \Delta x_{ij},
\end{eqnarray}
ensuring local conservation is given by
\begin{equation} \label{Conservation}
x_{i}^{\prime} + x_{j}^{\prime} = x_{i} + x_{j}.
\end{equation}
In the kinetic exchange simulations these rules are always maintained.

As mentioned above, a wide variety of kinetic exchange models have been proposed in the literature. In what follows, we briefly discuss several representative models and outline their underlying exchange mechanisms.

Several classical kinetic wealth-exchange models have been proposed to describe money transfer among agents under the constraint of total wealth conservation. For example; the Drăgulescu--Yakovenko model \cite{Dragulescu2000,yako01usa-income,Silva2005,Yakovenko_2009} assumes random pairwise exchanges according to
\begin{equation}
x_{i}^{\prime} = \varepsilon (x_{i} + x_{j}), \qquad 
x_{j}^{\prime} = (1 - \varepsilon)(x_{i} + x_{j}),
\end{equation}
where $\varepsilon \in [0,1]$ is a random number.  
This simple rule conserves total wealth as well in Eq.(\ref{Conservation}) and leads to an exponential steady-state distribution,
\begin{equation}
P(x) = \frac{1}{M} e^{-x/M},
\end{equation}
analogous to the Boltzmann--Gibbs law, where $M=<x_{i}>$. 

The Chakraborti--Chakrabarti model~\cite{Chakraborti_2000,Chakraborti_2002,Chatterjee_2007} extends this framework by introducing a uniform \emph{saving propensity} $\lambda$, representing the fraction of money each agent retains during a trade:
\begin{eqnarray}
x_{i}^{\prime} = \lambda x_{i} + \varepsilon (1-\lambda)(x_{i} + x_{j}),  \nonumber \\
x_{j}^{\prime} = \lambda x_{j} + (1-\varepsilon)(1-\lambda)(x_{i} + x_{j}).
\end{eqnarray}
where $\varepsilon \in [0,1]$ is a random number and $0 \le \lambda < 1$. In this interaction schema, money conservation is preserved in every interaction, as indicated in Eq.~(\ref{Conservation}). The model conserves total money at each step. When $\lambda = 0$, it reduces to the Drăgulescu--Yakovenko model, whereas $\lambda \rightarrow 1$ corresponds to a frozen economy.  
The stationary money distribution is Gamma-like:
\begin{equation} \label{CC-dist}
P(x) = \frac{1}{\Gamma(n)} \left( \frac{n}{M} \right)^{n}
x^{n-1} e^{-n x/M},
\end{equation}
where $n = \frac{1 + 2\lambda}{1 - \lambda}$ and $M = \langle x_{i} \rangle$. This model still conserves total wealth but narrows the distribution, yielding a Gamma-like stationary state that reduces inequality.  
A natural extension introduces \emph{heterogeneous saving propensities} $\lambda_i$ drawn from a given distribution, reflecting individual differences in saving behavior.  

The Chatterjee--Chakrabarti--Manna model ~\cite{Chatterjee_2003,Chatterjee_2004} explicitly incorporates such heterogeneity, with each agent $i$ characterized by its own $\lambda_{i}$.  
The microscopic rule becomes
\begin{eqnarray}
x_{i}^{\prime} = \lambda_{i} x_{i} +
\varepsilon \big[ (1 - \lambda_{i}) x_{i} + (1 - \lambda_{j}) x_{j} \big],  \nonumber \\
x_{j}^{\prime} = \lambda_{j} x_{j} +
(1 - \varepsilon) \big[ (1 - \lambda_{i}) x_{i} + (1 - \lambda_{j}) x_{j} \big],
\end{eqnarray}
where $\varepsilon \in [0,1]$ is a uniformly random distributed, and $\lambda_{i,j}$ represent the saving propensities of the interacting agents.  The interaction rule ensures that the total wealth of the system is conserved at every exchange, consistent with Eq.~(\ref{Conservation}). After sufficient interactions, the steady-state money distribution 
produce a Pareto-type steady-state tail,
\begin{equation}
P(x) \sim x^{-\nu}, \qquad \nu \approx 2,
\end{equation}
which captures wealth inequality emerging endogenously from diverse saving propensities.  

Several extensions of the basic kinetic-exchange framework have been proposed to incorporate risk-based exchanges, taxation, and stochastic investment processes. These generalized models are known to produce either exponential or Pareto-type steady-state distributions depending on the specific microscopic rules and noise terms \cite{Boghosian_2014,Cordier_2005}. Since our focus is on introducing a new trap–diffusion mechanism, we do not elaborate further on these variants here.

In contrast to conventional kinetic-exchange models \cite{Dragulescu2000,Chakraborti_2000,Chatterjee2005,Champernowne1953,
Garlaschelli2004,Atkinson2011,Yakovenko_2009,yako01usa-income,
Boghosian_2014,Cordier_2005,chak13book-income-inequal-wealth,
chak22income-distr-inequal,Silva2005,Chakraborti_2002,
Chatterjee_2003,Chatterjee_2004,Chatterjee_2007}, we propose a new one-dimensional exchange model with randomly distributed potential wells, which generates an exponential Gibbs--Boltzmann–like regime for small wealth ($x \lesssim T$) and a power-law Pareto regime for large wealth ($x \gg T$) where $T$ is the temperature. Kinetic Monte Carlo simulations demonstrate that the resulting wealth distributions exhibit clear Pareto-type tails and reveal that the Pareto exponent varies systematically with temperature in the range $0.5 < \nu(T) < 1.5$. The proposed model thus unifies exponential and Pareto regimes within a thermodynamically motivated framework and can be naturally extended to non-conservative open systems.

\section{Kinetic Trap-Diffusion Model} \label{KineticTDM}

\begin{figure}[h!]
    \centering
    \includegraphics[width=0.5\textwidth]{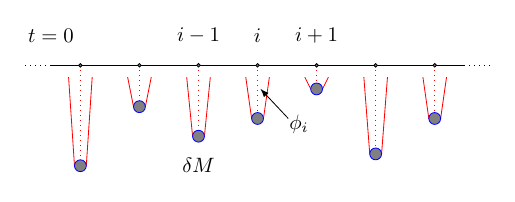}
    \caption{Schematic representation of the one-dimensional model, where the potential wells are randomly distributed along the lattice.}
    \label{Model}
\end{figure}

Similar to standard kinetic exchange approaches, the present model is formulated
as an agent-based framework.
A one-dimensional realization of the model is illustrated in
Fig.~\ref{Model}.
The system consists of a discrete lattice of length $L$, where each lattice
site is associated with a local potential well.
The money of each agent is represented by particles residing in the
corresponding potential well. Each agent $i$ is characterized by a fixed potential well depth $\phi_i$,
which defines the local energy landscape of the system.
The set of potential depths $\{\phi_i\}$ is randomly assigned at the initial
time $t=0$ and remains unchanged throughout the simulation.
In the present implementation, the values of $\phi_i$ are drawn from a uniform
distribution within the interval $[0,1]$.
This quenched heterogeneity generates a static disordered potential landscape
across the lattice. The total amount of money in the system is conserved during the dynamics, as
expressed by Eq.~(\ref{Constant_M}), ensuring that the model represents a closed
economic system.

In Fig.~\ref{Model}, $\delta M$ denotes the initial money \textit{amount} of each agent at $t=0$.
This value is taken to be constant in the simulations for simplicity.
The model can also be implemented with different initial values of $\delta M$.
However, this choice only affects the time required to reach the stationary distribution from a statistical point of view~\cite{Aydiner2023}.
For this reason, we fix the value of $\delta M$ throughout the present study.

The exchange mechanism operates as follows.
Money is represented by discrete units that occupy sites associated with local
potential wells of depth $\phi_i$.
Each money unit (particle) may attempt to hop between neighboring sites,
thereby modeling monetary exchanges between economic agents.
The energy cost associated with a proposed hop from site $i$ to a neighboring
site $j$ is determined by the difference in potential depths,
$\Delta U = \phi_i - \phi_j$.
Transitions toward deeper potential wells ($\Delta U \le 0$) are energetically
favorable and are always accepted, whereas transitions toward shallower wells
($\Delta U > 0$) require thermal activation.

The hopping dynamics is governed by a Metropolis acceptance rule.
A proposed transition from site $i$ to site $j$ is accepted with
probability
$P_{i \to j} = \min \left(1,\exp[-(\phi_i-\phi_j)/T]\right)$,
where $T$ denotes the effective temperature controlling the mobility
of money. In economic terms, $T$ may be interpreted as an effective
measure of money-exchange activity, since increasing $T$ facilitates
transitions across the potential landscape and thereby enhances the
frequency of successful money exchanges in the system.
At low temperatures, upward hops against the potential landscape are
strongly suppressed and money remains trapped in deep wells, whereas
at higher temperatures barrier-crossing events become increasingly
frequent.

The resulting microscopic exchange dynamics can be expressed in terms of the
stochastic difference equations
\begin{eqnarray}
x_i' &=& x_i + \Delta N_{ij}, \nonumber \\
x_j' &=& x_j - \Delta N_{ij},
\label{model-exchange}
\end{eqnarray}
where $\Delta N_{ij}$ denotes the amount of money exchanged between agents $i$
and $j$ at time $t$.
The exchange term $\Delta N_{ij}$ is activated only when the proposed hop
satisfies the Metropolis acceptance criterion, thereby explicitly coupling
monetary exchange to the local potential landscape and to temperature.

The particles diffuse on the lattice in discrete time intervals according to a kinetic random-walk dynamics.
After a sufficiently long period of steady-state evolution, the deepest potential well is, with high probability, the one that accumulates the largest amount of money.
However, the key parameter controlling the overall dynamics is the temperature.
If the temperature is very low, money exchange events may almost never occur;
conversely, if the temperature is extremely high, no agent can retain enough wealth to accumulate significant savings.
Between these two limiting regimes, the trading dynamics operates efficiently and leads to a well-defined stationary distribution of money.

Although the present model shares a similar exchange mechanism with the previous kinetic-exchange models,
it generalizes them by explicitly incorporating temperature as a control parameter.
Because of this temperature dependence, the Pareto exponent may vary as a function of $T$,
allowing the model—unlike conventional ones—to reproduce different Pareto exponents across distinct economic conditions.

\section{Monte Carlo Simulation} \label{MCSim}

The time evolution of the model schematically illustrated in Fig.~\ref{Model} is investigated using the Monte Carlo simulation implemented in Fortran. 
The numerical simulations are performed with the following parameter settings.
The system is defined on a one-dimensional lattice of length $L = 100$.
At the initial time $t = 0$, each lattice site is populated with
$\delta M = 10$ particles.
The elementary money exchange step is fixed to $\Delta N_{ij} = 1$ for simplicity.
The Monte Carlo simulation is evolved for $N = 5 \times 10^{4}$ time steps.
To ensure statistical reliability, all results are averaged over
$R = 1000$ independent realizations.
The simulations are repeated for different values of the temperature $T$,
while keeping all other parameters fixed.

The simulation procedure is implemented as follows: i)
First, a fixed value of the temperature $T$ is selected.
The dynamics is then simulated sequentially over the one-dimensional lattice,
starting from site $i = 1$ and proceeding site by site up to $i = L$.
ii) To illustrate the procedure, consider the dynamics associated with the first
site ($i=1$).
A single particle initially located in the potential well at site $i=1$ is
selected.
At the chosen temperature $T$, the particle attempts to hop to either the left
or the right neighboring well.
The proposed hop is accepted only if the particle successfully overcomes the
corresponding barrier, according to the Metropolis acceptance rule $P_{i \to j} = \min ( 1, \exp[ - (\phi_i - \phi_j)/ T ] )$.
This trial is repeated for a total of $N$ Monte Carlo steps, after which the
particle remains fixed at its final position and no further moves are allowed.
iii) The same simulation protocol is then applied independently to all other
particles initially located at site $i=1$.
iv) Once the simulation procedure for all particles at the first site is completed,
the algorithm proceeds to the next site ($i=2$), and the entire process is
repeated. This sequential procedure continues until all lattice sites
$i=1,\ldots,L$ have been processed, with periodic boundary
conditions imposed at the boundaries of the lattice.
v) After completing the simulations for all sites, each particle is counted at
its final position, corresponding to the potential well in which it ultimately
resides.
The resulting particle distribution defines a stationary wealth distribution
for a single realization of the one-dimensional system.
vi) Finally, to obtain reliable statistical results, the entire simulation procedure is
repeated over $R = 1000$ independent realizations.
The final stationary distribution is obtained by averaging over these
realizations.

\begin{figure}[h!] 
    \centering
\includegraphics[width=0.5\textwidth]{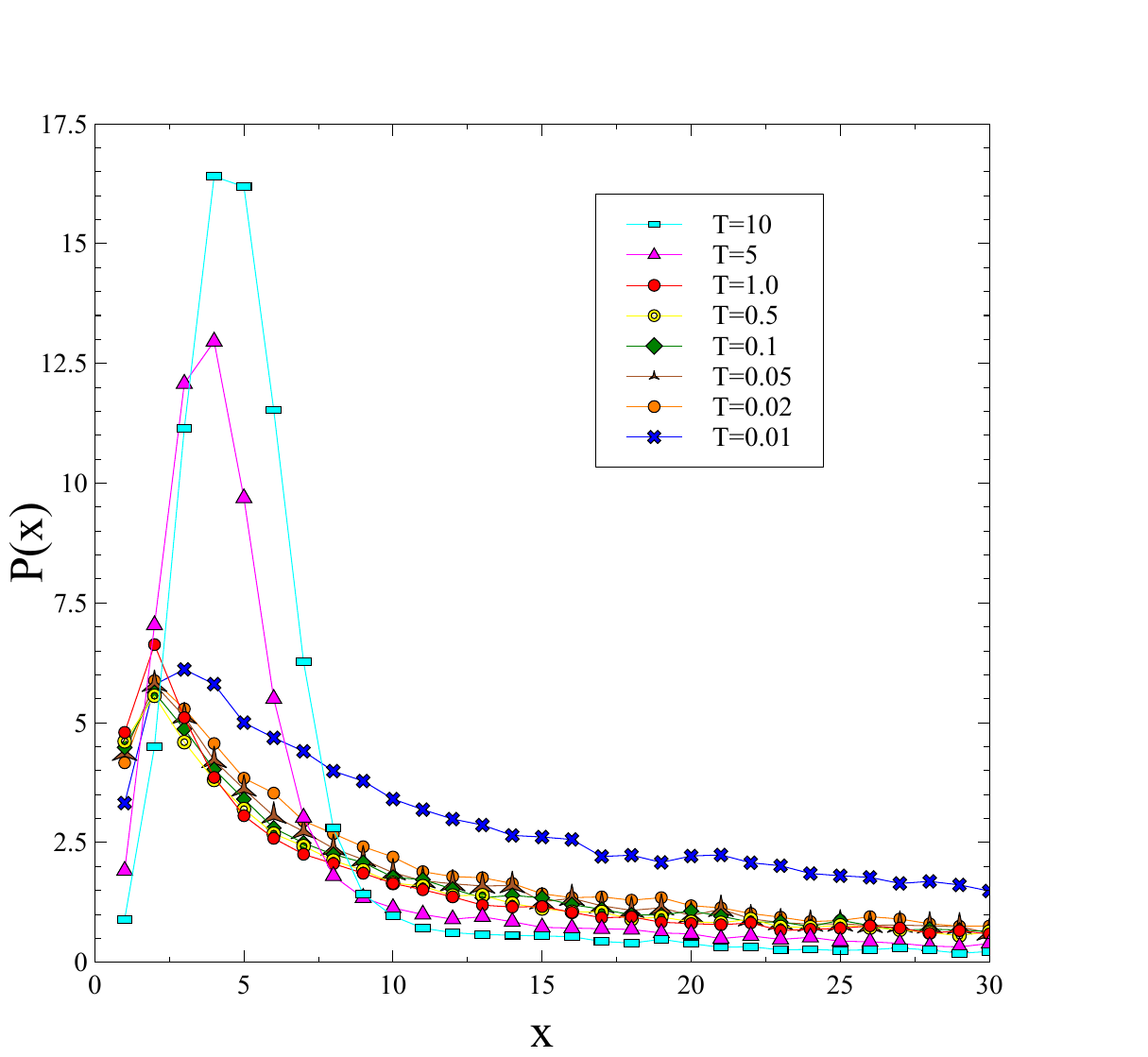}
    \caption{Pareto exponent for different temperatures and the corresponding money distribution profiles as a function of $x$.  
The results show that the money distributions follow an inverse power-law behavior, and the Pareto exponent $\nu(T)$ varies systematically with temperature.}
    \label{P(x)-dist}
\end{figure}

\begin{figure}[t] 
    \centering
\includegraphics[width=0.5\textwidth]{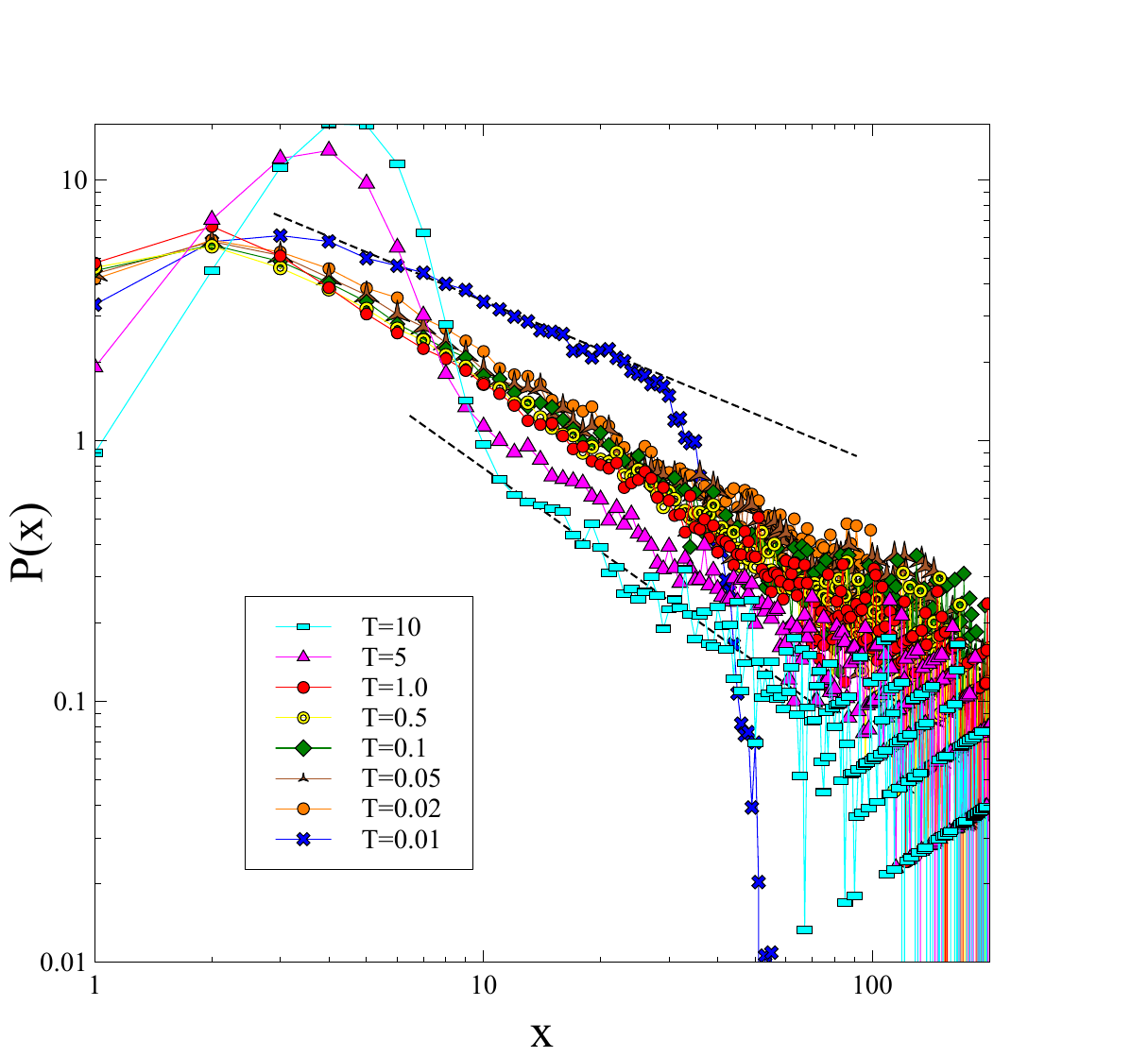}
    \caption{The money distribution $P(x)$ plotted on a log--log scale for different temperature values.  The tail of the distribution follows an inverse power-law behavior, clearly indicating a Pareto-type scaling. The fitted slopes (dashed lines) correspond to the Pareto exponents obtained for each temperature.}
    \label{log-P(x)-dist}
\end{figure}

The probability distribution \(P(x)\) obtained from the Monte Carlo simulation is presented in Fig.~\ref{P(x)-dist} as a function of \(x\) for various temperature values. As can be seen from the figure, \(P(x)\) exhibits a pronounced peak at small $x$ values, while the distribution gradually decreases and approaches zero for larger $x$.  The overall trend of the obtained probability distribution is qualitatively similar to those reported in previous studies. In particular, for high temperatures and small $x$ values, the distribution clearly exhibits a Gibbs–Boltzmann character. In this regime, the probability density is broad, smooth, and symmetric, indicating that the system possesses sufficiently high thermal energy to overcome potential differences. In this region, the distribution takes a nearly exponential form, $P(x) \sim e^{-x/x_{c}}$, where $x_{c}=x \sim 5$. In contrast, for low temperatures, the distribution becomes significantly narrower and forms a sharper peak. This behavior clearly shows that as the temperature decreases, the system becomes localized around the deepest minimum of the potential surface, and only low–energy states are predominantly occupied. Therefore, the figure reveals that while the classical Boltzmann–Gibbs behavior dominates in the high–temperature limit, the system transitions to a localized, low–entropy regime as the temperature decreases.

The probability distribution $P(x)$ for different temperature values is shown in Fig.~\ref{log-P(x)-dist} on a $\log$--$\log$ scale. As can be seen from the figure, the main tail of the distribution follows a power-law form remarkably well. Only for the very low temperature $T = 0.01$, a sudden deviation (blue square line) is observed beyond a certain value of $x$, which can be attributed to the loss of statistical accuracy due to limited sampling in the extreme tail region.  
Excluding this deviation and disregarding minor fluctuations at large $x$, the distributions can be convincingly described by a power-law dependence. The dashed lines in the lower (\textit{barhorz}- cyan color) and upper parts (\textit{cross}- blue color) in Fig.~\ref{log-P(x)-dist} represent the fitted curves, which clearly demonstrate the separation between the low- and high-temperature regimes. Accordingly, the probability distribution as a function of the temperature parameter can be expressed as  
\begin{equation} \label{temp-exp}
P(x;T) \sim x^{-\nu(T)} \, ,
\end{equation}
where the scaling exponent \(\nu(T)\) characterizes the temperature-dependent behavior of the tail.  

\begin{figure}[h!] 
    \centering
\includegraphics[width=0.45\textwidth]{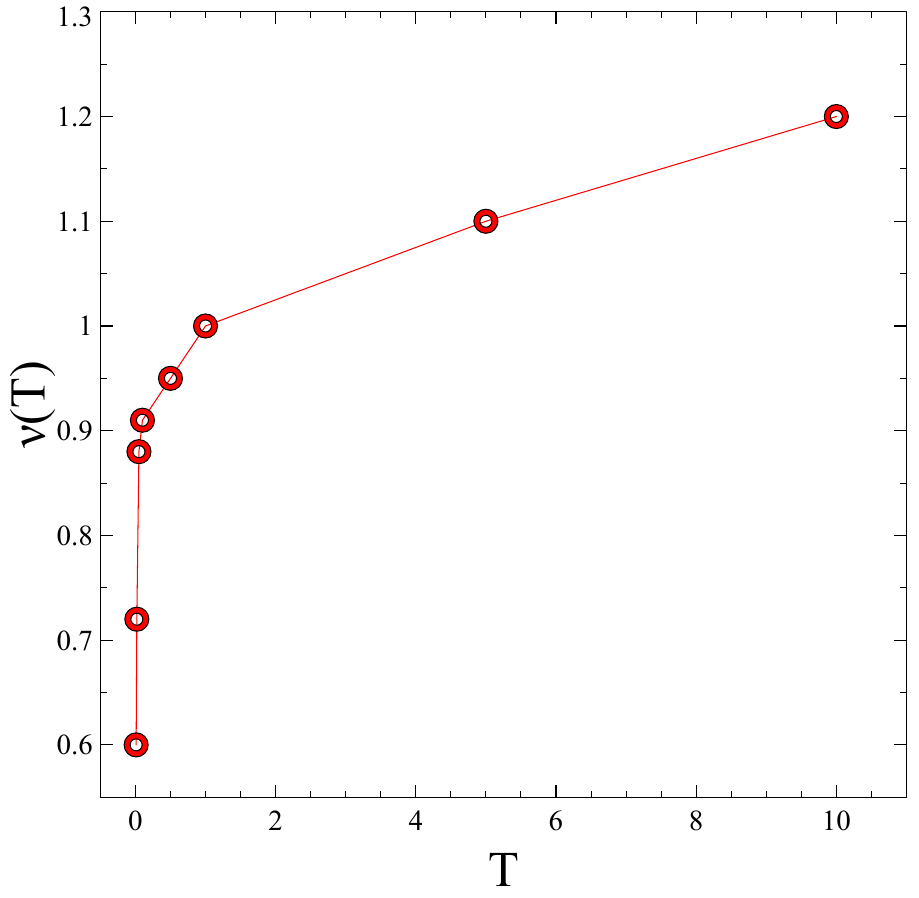}
    \caption{Values of the Pareto exponent $\nu(T)$ obtained for different temperature values.  
The exponent varies approximately within the range $0.5 < \nu(T) < 1.5$, clearly showing its dependence on temperature.}
    \label{exponent-temp}
\end{figure}

The temperature dependence of the exponent $\nu(T)$ appearing in Eq.~(\ref{temp-exp}) is presented in Fig.~\ref{exponent-temp}. As mentioned above, the temperature parameter $T$ determines the diffusive energy of the particles. For this analysis, the temperature values were chosen as $T = 0.01$, $0.02$, $0.05$, $0.1$, $0.5$, $1$, $5$, and $10$. The corresponding $\nu(T)$ values were obtained by fitting the tails of the distribution functions shown in Fig.~\ref{log-P(x)-dist}. As can be seen from the figure, the Pareto exponent $\nu(T)$ takes smaller values at low temperatures and increases gradually with increasing $T$. Within the simulation limits, the exponent varies approximately in the range $0.5 < \nu(T) < 1.5$. This result represents a parametric dependence, indicating that the relative temperature plays a crucial role in determining the scaling exponent. A higher numerical precision in the simulations would allow more accurate estimation of the upper and lower bounds of $\nu(T)$.

As can be inferred from Fig.~\ref{exponent-temp}, for sufficiently large values of the
temperature $T$, the Pareto exponent $\nu(T)$ may extrapolate toward larger values.
A clearer characterization of this high-temperature regime would require
simulations with extended parameter ranges, including larger lattice sizes,
longer simulation times, and a higher number of particles, which may
necessitate the use of high-performance computing resources.
Such extensions would allow a more detailed exploration of the accessible range
of $\nu(T)$.

It is important to emphasize, however, that the primary aim of the present work
is not to achieve particularly large values of the Pareto exponent.
Rather, the central objective is to demonstrate that the Pareto exponent can
vary systematically as a function of temperature. This temperature dependence
is particularly important because it establishes a direct connection between
the underlying trap--diffusion dynamics and the macroscopic form of the wealth
distribution. Since the Pareto exponent determines the heaviness of the
high-wealth tail, its variation with temperature reflects how changes in
trapping and monetary mobility modify the resulting degree of wealth
concentration. More importantly, Fig.~\ref{exponent-temp} demonstrates that different Pareto
exponents do not necessarily require different exchange rules or different
models: within the same trap--diffusion mechanism, changes in the effective
mobility of money are sufficient to generate different Pareto regimes.
Therefore, $\nu(T)$ provides a quantitative link between the microscopic
trapping and mobility processes and the resulting macroscopic wealth
distribution. 

The values of $\nu(T)$ obtained in this study are consistent with those reported
in empirical analyses of real economic systems
\cite{Pareto1897, gini21, elia12gini-soko, elia16harness,
mand60pareto-levy, bouc00conden-wealth, pers92, kato22islam-capital,
Gibrat1931, Dragulescu_2001, Fujiwara2003, gaba09rev-pareto-power-laws,
yako10njp, saez14inequal-science, levi14big-data, elia14rich-poor,
chak16inequality-kin-exch, gaba16inequality-well-cited,
shes17unequal-psycho, diff19glob-warm-pnas-inequal,
stoj22rgbm-income, kemp22, Souma_2002, Fujiwara_2003,
Atkinson_2011, Clementi_2005}.
Taken together, these results confirm that the proposed model not only
reproduces the Pareto law, but also yields exponent values compatible with
empirical observations, thereby supporting its consistency and robustness as
an alternative modeling approach.

\begin{figure}[h!] 
    \centering
\includegraphics[width=0.45\textwidth]{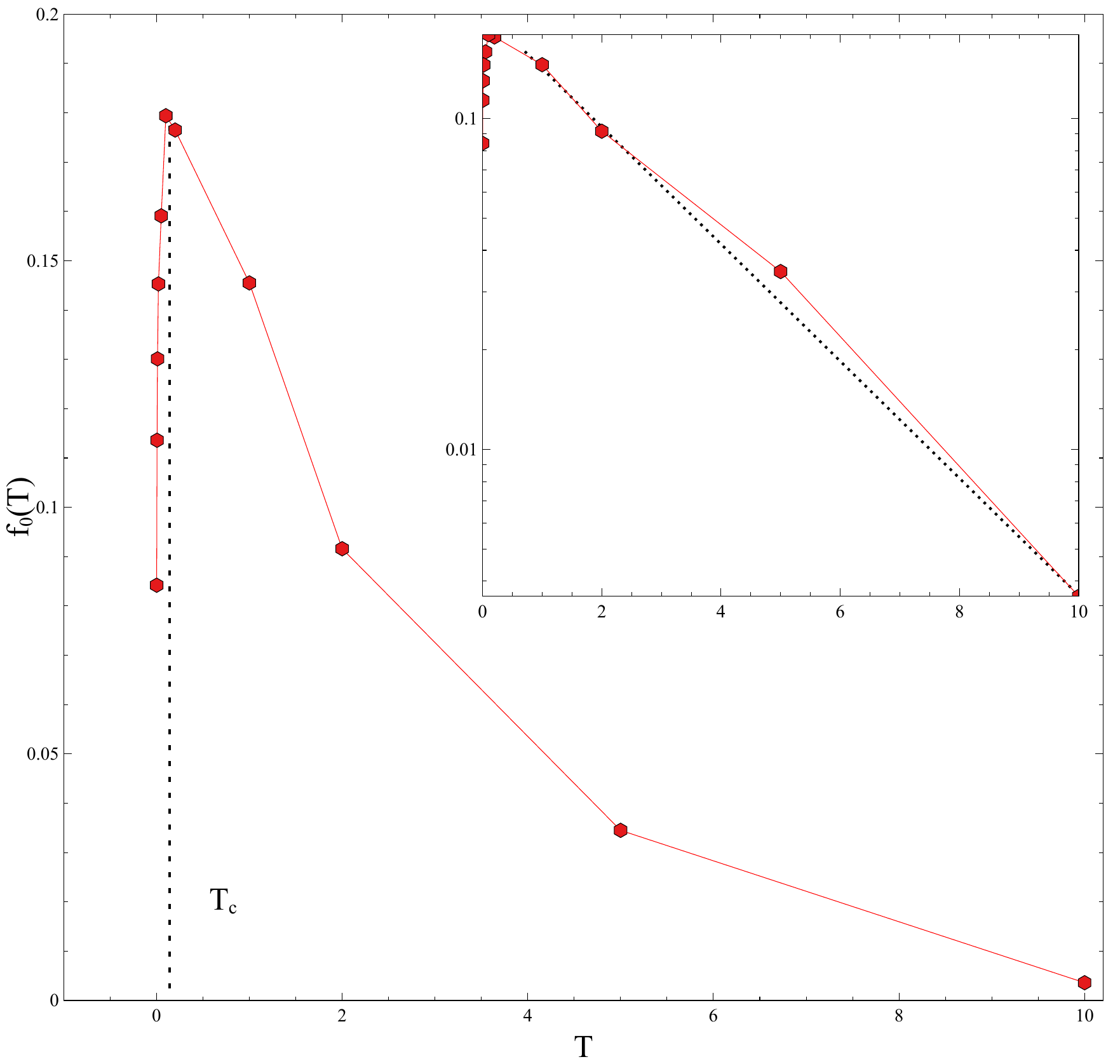}
      \caption{Vacancy fraction $f_{0}(T)$  as a function of temperature $T$. The dashed line indicates the critical temperature $T_{c} \simeq 0.05$.}
    \label{vacancy-temp}
\end{figure}

\begin{figure}[h!] 
    \centering
\includegraphics[width=0.45\textwidth]{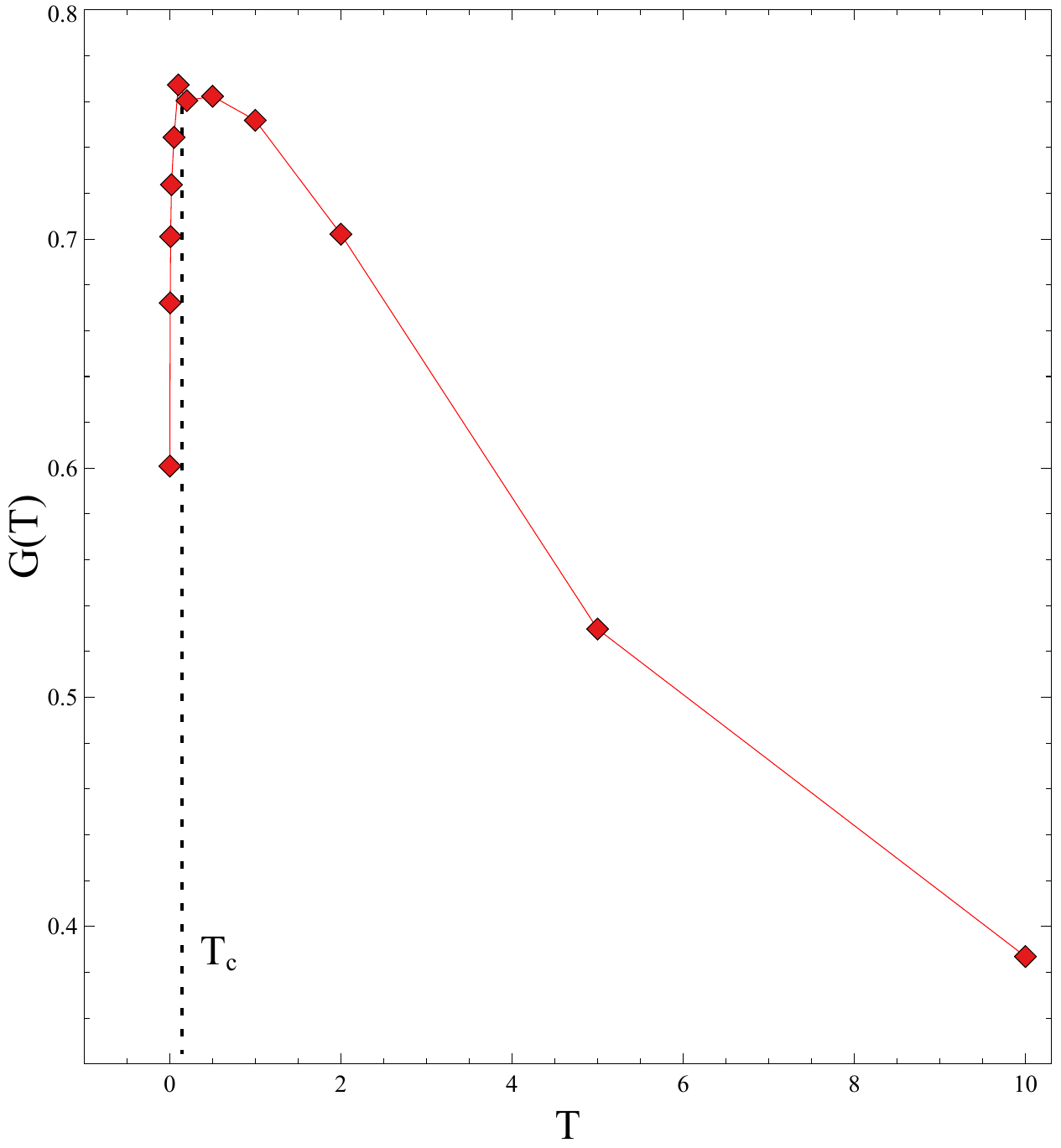}
    \caption{Gini index $G(T)$ as a function of temperature $T$. The dashed line indicates the critical temperature $T_{c} \simeq 0.05$.}
    \label{Gini-temp}
\end{figure}

To gain further insight into the underlying dynamics of the model,
we examined how the population of zero-wealth agents responds to
changes in the effective temperature. In Fig.~\ref{vacancy-temp},
the fraction of zero-wealth agents, $f_{0}(T)$, is shown as a function
of temperature. The simulations were performed for
$T=10,\ 5,\ 2,\ 1,\ 0.5,\ 0.2,\ 0.1,\ 0.05,\ 0.02,\ 0.01,$ and
$0.001$.

As can be seen from the figure, $f_{0}(T)$ exhibits a non-monotonic
temperature dependence. As the temperature decreases, the fraction
of zero-wealth agents initially increases and reaches a maximum in
the crossover region around $T\simeq0.05-0.1$. Upon further decreasing
the temperature, $f_{0}(T)$ decreases again.

For temperatures above the crossover region, the fraction of
zero-wealth agents decreases approximately exponentially with
increasing temperature, as demonstrated more clearly by the
semi-logarithmic representation shown in the inset. This behavior
indicates that enhanced money mobility suppresses the formation of
completely vacant sites by facilitating redistribution across the
lattice. At very low temperatures, in contrast, restricted mobility
and kinetic trapping become increasingly important and suppress the
complete depletion of individual sites. The maximum of $f_0(T)$
therefore reflects the competition between these two dynamical
mechanisms and indicates a crossover between a trapping-dominated
low-temperature regime and a mobility-dominated high-temperature
regime.

\begin{figure}[h!] 
    \centering
    \includegraphics[width=0.45\textwidth]{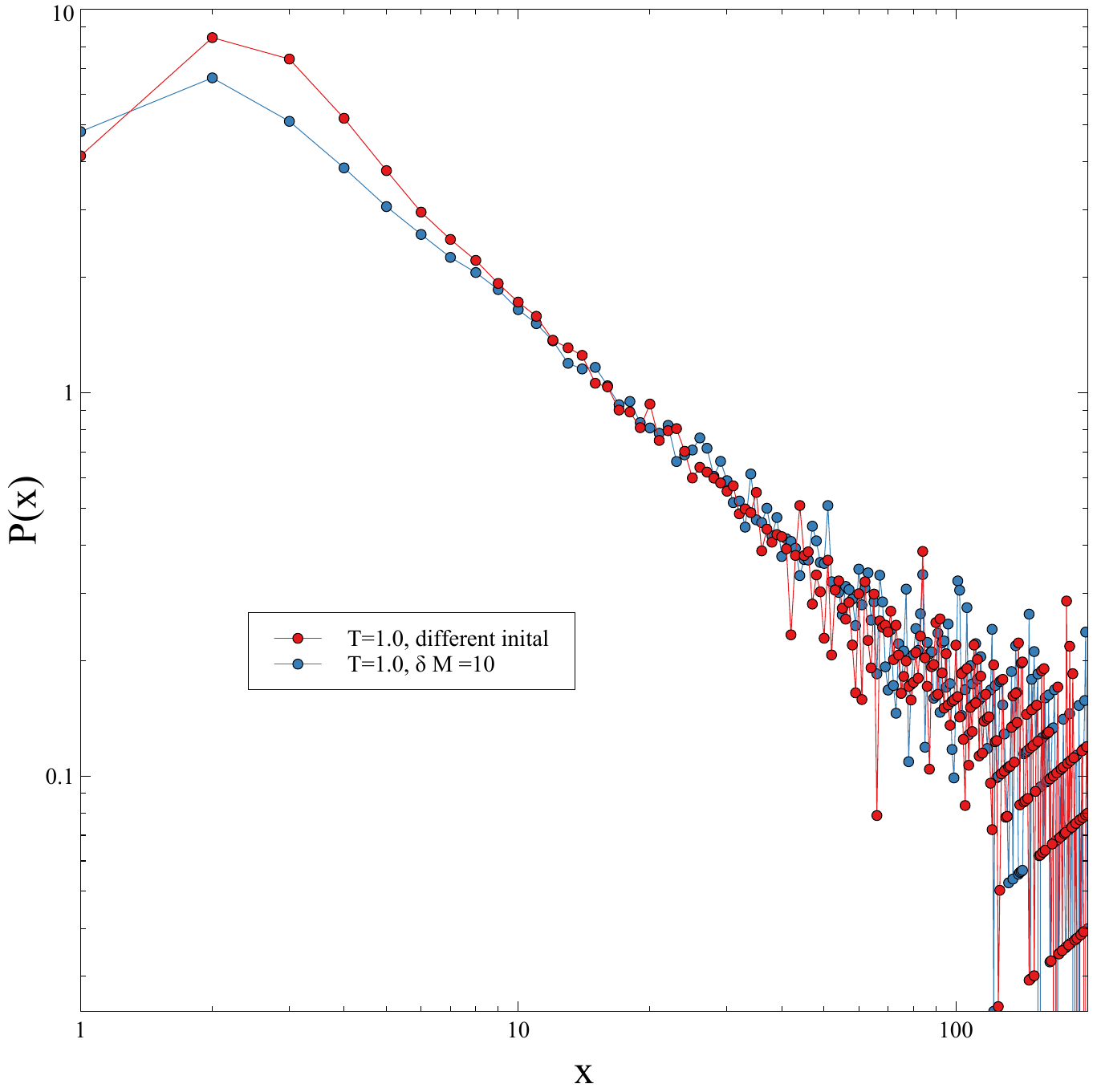}
    \caption{Stationary money distributions at $T=1.0$ for uniform and nonuniform initial conditions.}
    \label{x-irandom}
\end{figure}

The temperature dependence of the Gini index $G(T)$ is shown in
Fig.~\ref{Gini-temp}. Starting from the high-temperature regime,
$G(T)$ increases smoothly as the temperature is reduced and reaches
a maximum around the characteristic crossover temperature
$T_c\simeq0.05$. Upon further decreasing the temperature below
$T_c$, however, the Gini index decreases rapidly. Thus, the largest
wealth inequality is generated at an intermediate level of money
mobility rather than in either the very low- or high-temperature
limits.

This non-monotonic behavior can be understood from the underlying
exchange dynamics. In the very low-temperature regime, money
transfers are strongly suppressed and the system tends to remain
close to its initial wealth configuration. Since the system starts
from an approximately homogeneous distribution of money, the
suppression of exchange also limits the development of large wealth
differences, resulting in a decrease of $G(T)$ as $T$ approaches
zero. At intermediate temperatures, money becomes sufficiently
mobile to generate substantial redistribution, while the trapping
landscape still promotes localization and accumulation of money at
particular sites. The competition between these two effects produces
the maximum inequality around $T_c$. At still higher temperatures,
enhanced mobility facilitates barrier crossing and weakens the
localization of money, leading again to a reduction of the Gini
index.

In economic terms, this result suggests that inequality in the
present model is controlled by a competition between money-exchange
activity and wealth localization. Very weak exchange activity
restricts the development of wealth differences, whereas very strong
exchange activity prevents persistent localization of money. The
largest inequality therefore emerges at an intermediate exchange
activity where redistribution is active but trapping remains
effective. This non-monotonic dependence is a characteristic
consequence of the trap-diffusion dynamics of the present model.

Finally, to test the sensitivity of the stationary state to the initial
condition, we repeated the simulations using a strongly nonuniform
initial wealth configuration. Specifically, ten (10) lattice sites were
randomly selected and initially left empty, while the total money
$M=1000$ was randomly distributed among the remaining sites, so that
the total money was conserved. Fig.~\ref{x-irandom} compares the
stationary distributions obtained from the uniform and modified
initial conditions at $T=1.0$. The overall Pareto-like character of
the stationary distribution is preserved in both cases, although some
quantitative differences in the fitted Pareto exponent are observed.
This indicates that the emergence of the Pareto regime is robust
against the choice of the initial wealth configuration. This result is also consistent with our previous findings
in Ref.~\cite{Aydiner2023}, where the Pareto-like character of the
stationary wealth distribution was found to be robust against changes
in the initial conditions.


\section{Discussion} \label{Disc}

In the present work, we introduced a kinetic trap--diffusion model capable of
generating different Pareto exponents for money and income distributions.
Monte Carlo simulations demonstrate that the stationary wealth distribution
exhibits Pareto-type tails whose exponent varies systematically with the
temperature. Within this framework, temperature acts as a control parameter
that regulates monetary mobility and continuously tunes the statistical
properties of the wealth distribution.

The main advantage of the proposed trap--diffusion framework is its ability to
produce \emph{multiple} Pareto exponents as a function of a single control
parameter. At the same time, the model retains important connections with
classical kinetic exchange models
\cite{Dragulescu2000,Chakraborti_2000,Chatterjee2005,Champernowne1953,
Garlaschelli2004,Atkinson2011,Yakovenko_2009,yako01usa-income,
Boghosian_2014,Cordier_2005,chak13book-income-inequal-wealth,
chak22income-distr-inequal,Silva2005,Chakraborti_2002,
Chatterjee_2003,Chatterjee_2004,Chatterjee_2007}.
In conventional kinetic exchange models, the fraction of money participating
in transactions is often regulated by an explicitly introduced saving
propensity, $\lambda$, such that only the fraction $(1-\lambda)$ is available
for exchange. In the present model, no explicit saving propensity is imposed.
Instead, a conceptually similar restriction on exchange emerges dynamically
from the random trapping landscape: money located in sufficiently deep traps
may remain temporarily immobilized and therefore does not effectively
participate in exchange events. Thus, although saving and trapping have
different microscopic origins, both mechanisms restrict the effective
participation of money in the exchange process. The quenched heterogeneity
introduced through randomly distributed trap depths may therefore be viewed
as a dynamical counterpart of the heterogeneous agent properties commonly
considered in kinetic-exchange settings.

On the other hand, it should be emphasized that the effective temperature $T$ employed in the
present trap--diffusion model is not identical to the  \textit{money temperature}
often introduced in conventional kinetic exchange models through the
Gibbs--Boltzmann distribution, where the corresponding scale is typically
related to the average money per agent, $T_m=\langle m\rangle=M/N$. In the
present conservative system, $M$ and $N$ are fixed and hence $\langle
m\rangle$ remains constant. The temperature $T$ instead enters the
thermally activated transition probabilities associated with the random
trapping landscape. It therefore controls the escape rates, waiting times,
and effective mobility of money. Consequently, $T$ influences the stationary
money distribution indirectly through the exchange dynamics rather than
simply setting the exponential scale of a Gibbs--Boltzmann distribution.
This distinction is essential for understanding why the shape of the
stationary distribution, and in particular the Pareto exponent $\nu(T)$, can
change with $T$ even though the average amount of money in the closed system
remains fixed.

A key distinction of the present approach is that monetary mobility is governed
by a continuous-time random walk (CTRW)
\cite{Montroll1965,ScherMontroll1975,Klafter1987,Metzler2000} on a disordered
potential landscape. Due to site-dependent waiting times, the CTRW dynamics
constrains diffusion such that, at each update, only a fraction of agents
effectively participates in exchange events, while others remain temporarily
localized in deep traps \cite{Bouchaud1992,Monthus1996}. The temperature $T$
controls the activation rates and waiting-time statistics of this process,
thereby regulating how easily money escapes from local traps and redistributes
across the system. As a consequence, the stationary money distribution yields
power-law (Pareto-type) tails,
$P(x)\sim x^{-\nu(T)}$, with a temperature-dependent exponent $\nu(T)$.

The additional numerical results provide further insight into the dynamical
origin of these stationary wealth distributions. In particular, the vacancy
fraction $f_0(T)$ and the Gini index $G(T)$ exhibit a non-monotonic response
to the effective temperature. Remarkably, both quantities show their
characteristic change of behavior in approximately the same temperature
region, $T_c\simeq0.05$. Since these observables characterize different
aspects of the wealth distribution, their common temperature scale suggests
that the observed crossover is not specific to a single statistical measure,
but reflects a characteristic dynamical feature of the trap--diffusion
process.

This behavior can be understood in terms of the competition between trapping
and thermally activated money mobility. At very low temperatures, transitions
across the random potential landscape are strongly suppressed. Consequently,
the system explores only a limited part of the accessible configuration space
and large-scale wealth redistribution is inhibited. As the temperature
increases, money becomes sufficiently mobile to explore the heterogeneous
trapping landscape, allowing localization and accumulation at particular
sites. This process enhances both the population of zero-wealth agents and
the overall wealth inequality. The simultaneous maxima of $f_0(T)$ and
$G(T)$ around the crossover region suggest that the competition between
mobility and localization becomes particularly important in this regime.
At still higher temperatures, barrier crossing becomes increasingly frequent,
the influence of individual traps is weakened, and money is more efficiently
redistributed over the lattice. As a consequence, both the vacancy fraction
and the Gini index decrease.

These results also provide a possible economic interpretation of the effective
temperature. Within the present model, $T$ should not be identified with a
conventional thermodynamic temperature or with a specific macroeconomic
variable. Rather, it may be regarded as an effective measure of money-exchange
activity or economic mobility. From this perspective, the relation between
exchange activity and inequality need not be monotonic. Very weak exchange
activity restricts the development of large wealth differences, whereas very
strong mobility prevents persistent localization of money. The largest
inequality therefore emerges at an intermediate exchange activity, where
redistribution is sufficiently active to generate substantial wealth
heterogeneity while trapping remains sufficiently effective to sustain
localization.

An important implication of the model is that the temperature-dependent
Pareto exponent, the fraction of zero-wealth agents, and the Gini index all
emerge from the same underlying trap--diffusion dynamics rather than from
separate phenomenological mechanisms. Thus, qualitatively different
wealth-distribution regimes can arise within the same exchange dynamics
simply through changes in the effective mobility of money. This interpretation
should, however, be understood as a consequence of the present kinetic model
rather than as a direct prediction for a particular real economy.

The simulations performed with a strongly nonuniform initial wealth
configuration provide an additional robustness test. The Pareto-like
character of the stationary distribution survives a substantial modification
of the initial state. Although quantitative changes in the fitted Pareto
exponent are observed, the persistence of the Pareto regime indicates that
its emergence is not tied to the initially uniform distribution of money.
Together with our previous results~\cite{Aydiner2023}, this supports the
robustness of the basic mechanism responsible for the formation of the
stationary heavy-tailed wealth distribution.

Finally, the present study focuses on conservative exchange dynamics in order
to isolate the role of temperature in a controlled setting. Real economic
systems are, however, inherently non-conservative, involving money creation,
dissipation, taxation, debt, and external flows. Consistent with our earlier
investigations of open agent-based systems
\cite{Aydiner2018,Aydiner2019,Aydiner2023}, the present framework can be
naturally extended to non-equilibrium open economic systems. Such extensions
may incorporate heterogeneous initial conditions, debt
\cite{bogh19negative-wealth}, and wealth-resetting shocks associated with
financial crises and crashes
\cite{sorn03crash-book,bouc03bubble-crash-agents,cher21gbm,
vino22reset-gbm,cher22reset-gbm,trif21rgbm}. At the opposite limit of very slow dynamics, stochastic resetting
has also been investigated for ultraslow diffusion processes,
revealing non-equilibrium stationary states and characteristic
first-passage properties~\cite{Liang_2025}. These extensions may provide
a useful route for testing how the trap--diffusion mechanism operates in more
realistic economic environments.

\section{Conclusion} \label{Conc}

In this work, we introduced a one-dimensional kinetic trap--diffusion
model for money exchange in which the effective temperature controls
monetary mobility on a heterogeneous potential landscape. Monte Carlo
simulations show that the stationary money distribution exhibits a
Gibbs--Boltzmann--like regime at small money values and Pareto-type
power-law tails at large money values, with a Pareto exponent that
varies systematically with temperature.

Beyond the temperature-dependent Pareto exponent, the vacancy fraction
$f_0(T)$ and the Gini index $G(T)$ reveal a non-monotonic dependence
on temperature and exhibit a characteristic crossover around
$T_c\simeq0.05$. These results indicate that the stationary wealth
distribution is governed by the competition between money mobility
and trapping on the heterogeneous landscape. The persistence of the
Pareto-like regime under strongly nonuniform initial conditions further
supports the robustness of the underlying mechanism. Overall, the
model provides a minimal dynamical framework in which different
wealth-distribution and inequality regimes emerge from the same
trap--diffusion dynamics through changes in a single control parameter.

\section*{Acknowledgments}

The author is grateful to Professor Igor M. Sokolov for many helpful and insightful discussions that contributed to the conception and development of this work.
The author also thanks Prof. Paul J. Steinhardt for his hospitality during his stay at Princeton University. The author acknowledges Princeton University for providing a stimulating academic environment during this period.
This study was partially supported by T\"{U}B\.{I}TAK under the 2219 Project. The author also sincerely thanks the reviewers for their valuable and critical comments, which have significantly improved the manuscript.

\bibliographystyle{apsrev4-2}
\bibliography{Pareto}

\end{document}